\documentclass[5p,sort,compress]{elsarticle}
\makeatletter
\def\@preprint{\@empty}
\newcommand{\preprint}[1]{\gdef\@preprint{\hfill #1}}
\long\def\MaketitleBox{%
  \resetTitleCounters
  \@preprint\par
  \def\baselinestretch{1}%
  \begin{\elsarticletitlealign}%
   \def\baselinestretch{1}%
    \Large\@title\par\vskip18pt
  \ifdoubleblind
    \vspace*{2pc}
  \else
    \normalsize\elsauthors\par\vskip10pt
    \footnotesize\itshape\elsaddress\par\vskip36pt
  \fi
    \end{\elsarticletitlealign}%
}
\makeatother
\usepackage{graphicx}
\usepackage{color}
\graphicspath{{figures/}{fig/}}
\usepackage{caption}
\usepackage{subcaption}
\usepackage{amsmath}
\usepackage{amssymb}
\usepackage{bm}
\usepackage{slashed}
\usepackage{epsfig}
\usepackage{amsfonts}
\usepackage{hyperref}
\usepackage{bbm}
\usepackage{textcomp}
\usepackage{color}
\usepackage[ruled,linesnumbered]{algorithm2e}
\usepackage[T1]{fontenc}
\newcommand{\sect}[1]{{\it #1.  }}
\begin{document}
\title{Reclassifying Feynman Integrals as Special Functions}

\author[aff1]{Zhi-Feng Liu\corref{cor1}}
\ead{zfliu1013@163.com}

\author[aff2,aff3]{Yan-Qing Ma\corref{cor1}}
\ead{yqma@pku.edu.cn}

\author[aff4]{Chen-Yu Wang\corref{cor1}}
\ead{cywang@mpp.mpg.de}

\cortext[cor1]{Corresponding authors.}

\affiliation[aff1]{
    organization = {Zhejiang Institute of Modern Physics, School of Physics, Zhejiang University},
    city = {Hangzhou 310027},
    country = {China}
}
\affiliation[aff2]{
    organization = {School of Physics, Peking University},
    city = {Beijing 100871},
    country = {China}
}
\affiliation[aff3]{
    organization = {Center for High Energy Physics, Peking University},
    city = {Beijing 100871},
    country = {China}
}
\affiliation[aff4]{
    organization = {Max-Planck-Institut f{\"u}r Physik},
    city = {Garching 85748},
    country = {Germany}
}

\journal{Science Bulletin}

\preprint{MPP-2023-265}

\maketitle

Feynman integrals (FIs) serve as fundamental components of perturbative quantum
field theory. The study of FIs is important for both exploring the mysteries of
quantum field theories and their phenomenological applications, particularly in particle physics. A significant amount of effort has been devoted to analytically calculate FIs, with the goal of expressing them as linear combinations of special functions. However, this approach encounters significant challenges, given the inclusion of relatively less explored special functions, such as those defined on elliptic curves~\cite{Laporta:2004rb} and Calabi-Yau manifolds~\cite{Klemm:2019dbm}.
Alternatively, an effective strategy for addressing FIs involves employing numerical differential equations~\cite{Remiddi:1997ny} and the auxiliary mass flow (AMFlow) method~\cite{Liu:2017jxz,Liu:2020kpc,Liu:2021wks,Liu:2022mfb,Liu:2022tji}. This approach, in principle, allows for the precise computation of any FI. Consequently, a shift in perspective can be adopted by considering FIs as a new
class of special functions, the study of which can also help to understand the aforementioned
relatively less explored special functions. To advance this line of inquiry,
there is a need for a more comprehensive investigation of the properties of
FIs.

\sect{Space of special functions}
The inquiry into the types of transcendental numbers or special functions that might manifest in scattering amplitudes has intrigued researchers for quite some time. In recent years, the differential equation method~\cite{Remiddi:1997ny} has emerged as a paramount tool for addressing this question.

A $L$-loop FI with $N$ propagators is usually defined as
\begin{equation}\label{eq:FIdef}
I_{\vec{\nu}}
=
\int
\prod_{i = 1}^{L} \frac{\mathrm{d}^{d} l_{i}}{\text{i} \pi^{d / 2}}
\prod_{a = 1}^{N} \frac{1}{{\cal D}_{a}^{\nu_{a}}}
,
\end{equation}
where $d = 4 - 2 \epsilon$ is the spacetime dimension, $l_{i}$ are loop momenta to be integrated out, ${\cal D}_{a}$ are inverse propagators which are polynomials of masses and
scalar products of momenta, and $\nu_{a}$ are integers. A FI
is an analytic function of kinematic invariants $s_{i}$, which can be either
scalar products of external momenta or masses in the propagators. It has been established that, given a set ${\cal D}_{a}$, any  $I_{\vec{\nu}}$ can be expressed as a linear combination of a finite set of basis integrals, denoted as $I_{i}$ and referred to as master integrals, through the FI reduction technique~\cite{Chetyrkin:1981qh,Smirnov:2010hn}:
\begin{equation}
	I_{\vec{\nu}}
	=
	\sum_{i = 1}^{M}
	c_{\vec{\nu},i} \, I_{i}
	,
\end{equation}
where $M$ is an integer, and $c_{\vec{\nu},i}$ are rational functions of $\epsilon$ and $\vec{s}$. Upon differentiating $I_{i}$ with respect to kinematic invariants and subsequently reducing back to the same basis, it becomes evident that the FIs $I_{i}$ satisfy a system of differential equations:
\begin{equation}\label{eq:DEs}
\frac{\partial}{\partial s_{i}} \boldsymbol{I}(\epsilon, \vec{s}~)
=
\boldsymbol{A}_i(\epsilon, \vec{s}~) \boldsymbol{I}(\epsilon, \vec{s}~)
,
\end{equation}
where $\boldsymbol{A}_i$ represents matrices with elements expressed as rational functions of both $\epsilon$ and $\vec{s}$.

In certain instances, through the selection of appropriate master integrals $\boldsymbol{I}'$, the differential equations can be transformed into the so-called ``canonical form''~\cite{Henn:2013pwa}, where the
dependence on $\epsilon$ factorizes
\begin{equation}
\frac{\partial}{\partial s_{i}} \boldsymbol{I}'(\epsilon, \vec{s}~)
=
\epsilon
\boldsymbol{A}'_i(\vec{s}~) \boldsymbol{I}'(\epsilon, \vec{s}~)
.
\end{equation}
The resultant equations often assume a highly compact form, and the analytical solution at each order of the $\epsilon$ expansion can be readily expressed using iterated integrals. In the simplest scenario, these iterated integrals can be characterized as multiple polylogarithms~\cite{2011arXiv1105.2076G}. Leveraging the intricate algebraic structures inherent in multiple polylogarithms, such as the symbol technique~\cite{Goncharov:2010jf}, we can significantly simplify the expression. Alternatively, one can directly formulate an ansatz for the final result of the FIs and subsequently determine the unknown coefficients by incorporating additional knowledge about the FIs. This strategic approach has proven instrumental in resolving numerous cutting-edge phenomenological problems in recent years.

In a more general context, the system of first-order differential equations can be reformulated as a differential operator acting on one of the master integrals. The subsequent factorization of this differential operator into a product of simpler counterparts provides insights into the special functions involved in the solution~\cite{Adams:2017tga}. If all factors are first-order, one would anticipate that the solution can be expressed in terms of multiple polylogarithms. Conversely, encountering an irreducible second-order differential operator suggests that the iterated integral solution will encompass an elliptic curve. From this standpoint, the space of special functions can assume arbitrary complexity in the absence of a strong constraint on the form of the differential operator.\footnote{Based on the resurgence theory~\cite{ecalle1981fonctions},
the perturbative series can eventually recover all nonperturbative information
of quantum field theory. Therefore, it is natural to expect that FIs with more
and more loops will be extremely complicated.} Indeed, in addition to multiple polylogarithms and elliptic curves, researchers have identified differential operators associated with generalized hypergeometric functions~\cite{Ablinger:2017bjx}, algebraic curves with higher genus, and more intricate geometric objects such as Calabi-Yau manifolds~\cite{Klemm:2019dbm}. Exploring special functions beyond multiple polylogarithms presents not only a current challenge in leveraging the symbol technique but also difficulty in obtaining numerical results at specific kinematic points. This area of investigation is currently a focal point of active research.

Besides exploring special functions within the $\epsilon$ expansion of FIs, another avenue of inquiry involves the comprehensive classification of dimensionally regulated FIs. This endeavor encompasses efforts to categorize FIs using GKZ systems~\cite{IMGel'fand_1992}. While constituting a top-down approach to understanding FIs, certain properties of GKZ systems remain underexplored, leaving numerous open questions. Challenges include the construction of the system and its expansion in terms of $\epsilon$. Additional research is imperative before it can effectively support state-of-the-art computations.

\sect{Semi-analytical computation}
The AMFlow method, initially proposed in 2017~\cite{Liu:2017jxz}, has undergone significant development in recent years~\cite{Liu:2020kpc,Liu:2021wks,Liu:2022mfb}. It now possesses the capability to calculate any FI to arbitrary precision, providing ample computational power.  The method has been implemented in the computer program~\cite{Liu:2022tji}, which can calculate FIs fully automatically.
In this methodology, it is customary to substitute one inverse propagator ${\cal D}_{a}$ in Eq.~\eqref{eq:FIdef} with ${\cal D}_{a} + \text{i} \eta$. Subsequently, we establish differential equations for the modified master integrals $\widetilde{\boldsymbol{I}}$ with
respect to the ``auxiliary mass'' term $\eta$,
\begin{equation}
\frac{\partial}{\partial \eta} \widetilde{\boldsymbol{I}}(\epsilon, \vec{s}~,\eta)
=
\widetilde{\boldsymbol{A}}(\epsilon, \vec{s}~,\eta) \widetilde{\boldsymbol{I}}(\epsilon, \vec{s}~,\eta)
,
\end{equation}
which can be achieved by using FI reduction technique.  If a boundary condition for the differential equations is available, the original FIs ${\boldsymbol{I}}$ can then be obtained by numerically solving these ordinary differential equations.

Fortunately, the boundary condition at $\eta \to \infty$ is always attainable. In this limit, the modified master integrals transform into linear combinations of three types of FIs:
The first type involves removing the inverse propagator ${\cal D}_{a}$, simplifying the calculation into a recursive problem with the same structure;
The second type consists of factorized FIs, which are notably simpler and amenable to recursive solutions; And the last type comprises one-mass vacuum FIs, exhibiting no external momentum dependence. To calculate the last type of FIs, we relate them to FIs with two external legs but having one less loop, which can again be calculated by using the above AMFlow method~\cite{Liu:2022mfb}.
By applying this method recursively, any FI can be comprehensively determined, relying solely on input from FI reduction that entails purely linear algebraic operations.

The AMFlow method offers a convenient means of evaluating FIs at any kinematic point. For practical applications, one can systematically cover the entire kinematic space by assessing each individual point using the AMFlow method. This involves first fixing the kinematic variables and then progressing along the auxiliary parameter direction to derive the integral value. Although this approach is straightforward to implement and parallelize, it overlooks the fact that the values of FIs at two nearby kinematic points often exhibit smooth changes, leading to suboptimal efficiency.

\begin{figure}[htb]
    \centering
        \includegraphics[width=0.23\textwidth]{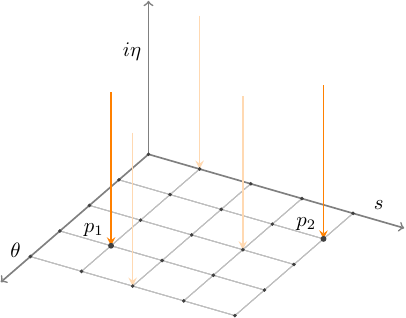}
        \includegraphics[width=0.23\textwidth]{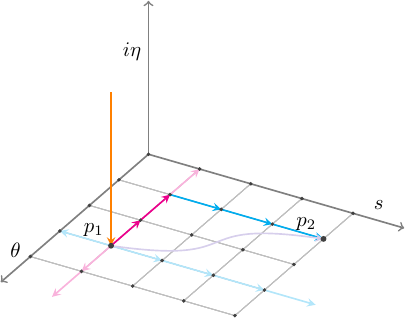}
    \caption{(Color online) Comparison of different evaluation strategies. Left: Using {AMFlow} point by point; right: Using {AMFlow} to provide a boundary condition and then using differential equations with respect to kinematic variables.}
    \label{fig:kinematic_space}
\end{figure}

The dynamics of how FIs evolve in the kinematic space is inherently encoded in the differential equations concerning the kinematic variables that define the space. Hence, once the integral value at one kinematic point is known, it proves more convenient to navigate directly between points in the kinematic space rather than traversing the auxiliary parameter dimension every time.

The distinction between the two approaches is evident in Fig.~\ref{fig:kinematic_space}. In this illustration, the kinematic space is characterized by the center-of-mass energy $s$ and scattering angle $\theta$, with the auxiliary mass parameter denoted as $\eta$ in the context of the AMFlow method. In the first approach, each point in the kinematic space is evaluated along the $\eta$ direction. In the second approach, however, point $p_{1}$ is initially assessed using AMFlow, and subsequently, point $p_{2}$ is evaluated using the differential equation along the $\theta$ direction, followed by the $s$ direction. This flexible approach also allows movement along any curve in the kinematic space, provided that the correct analytic continuation is applied.

Solving differential equations in kinematic space introduces alternative methods for evaluating points. In a one-dimensional kinematic space, covering it with a set of series solutions facilitates efficient FI evaluation anywhere within the space. For lower-dimensional kinematic spaces, constructing a grid of series solutions is feasible. However, in higher-dimensional kinematic spaces, the rapid growth in grid size suggests that employing importance sampling would be a more efficient strategy. Consequently, by integrating the AMFlow method with numerical differential equations concerning kinematic variables, FIs can be systematically and efficiently calculated.

\sect{Feynman integrals as special functions}
Although FIs cannot be
expressed as well-studied special functions, they can be calculated
systematically and efficiently using the AMFlow method in combination
with differential equations with respect to kinematic variables, as we have discussed above.
Therefore, it is constructive to define FIs as a new class of special functions
(or transcendental numbers if there is no kinematic variable in FIs). The
so-called special functions often exhibit the following traits:
\begin{enumerate}
    \item[(1)] Having both integral and differential representations;
    \item[(2)] Well-studied asymptotic behavior around singularities and branch
        cuts;
    \item[(3)] Availability of series expansions everywhere;
    \item[(4)] Satisfying certain algebraic relations among them.
\end{enumerate}
These traits facilitate the exploration of global and local properties of the
function space, and also provide efficient evaluation methods. We will show that
FIs can indeed satisfy all these points, although better strategies for many
aspects are still needed.

(1) FIs have integral representations by definition, and master integrals, which
are bases of FIs, have closed differential equations with respect to kinematic
variables. Considering also that boundary conditions can always be obtained
through the AMFlow method, the first condition is satisfied.
Nevertheless, it should be pointed out that there is no differential equation
with respect to $\epsilon$, therefore $\epsilon$ should be thought as a
parameter instead of an argument for the special functions. By expanding
$\epsilon$ around the origin,
\begin{equation}\label{eq:Exp}
\boldsymbol{I}(\epsilon, \vec{s}~)
= \sum_i \boldsymbol{I}^{(i)}(\vec{s}~) \epsilon^i
,
\end{equation}
the coefficients can also be defined as special functions, and original FIs serve as
generating functions for these special functions.

(2) The singularities of FIs are determined by Landau equations, but these
equations are usually very hard to solve. Alternatively, singularities can be
identified as a subset of poles in differential equations, say $A_i$ in
Eq.~\eqref{eq:DEs}. Spurious poles can be ruled out by inspecting their
monodromy groups, as those groups associated with spurious poles act trivially on master
integrals. Branch cuts are usually identified by studying the Feynman
prescription $\text{i} 0^+$ in propagators, although this is not always an easy way
to tackle the problem, especially when some of the propagators are replaced by
delta functions. The bottom line is that we can use AMFlow to compute
some points around a branch point, so that we can fully determine the asymptotic
expansion at the point and thus determine the corresponding branch cut.

(3) With differential representation, boundary conditions and singular structures
discussed above, we can obtain series expansion at any desired point by analytic
continuation.

(4) Relations among FIs have been explored by the means of integration by parts
identities. However, it is not yet clear whether integration by parts identities exhausted
all possibilities. Furthermore, there can be more relations among coefficients
after $\epsilon$ expansion shown in Eq.~\eqref{eq:Exp}, as hinted by the study
of multiple polylogarithms. Unfortunately, there is currently no efficient way
to identify these relations. As AMFlow can compute the coefficients to
high precision, we can at least explore their relations by PSLQ algorithm.

\sect{Outlook}
We have demonstrated that, thanks to the input from AMFlow, it is
feasible to define FIs as a novel class of special functions. Several crucial
avenues for further exploration emerge in this direction:
\begin{itemize}
    \item[(1)] The development of an efficient technique for determining
    singularities and branch cuts is essential.
    \item[(2)] A refinement of the methodology for choosing master integrals is
        imperative. This process should be sufficiently general to apply across
        diverse cases, resulting in a simple matrix $\boldsymbol{A}_i(\epsilon,
        \vec{s})$ in Eq.~\eqref{eq:DEs}, conducive to efficient numerical
        computation.
    \item[(3)] A systematic approach to exploring relations between coefficients
        in the $\epsilon$ expansion is warranted.
\end{itemize}

Furthermore, exploring connections between FIs and established special functions
can facilitate a deeper understanding of the latter unique mathematical entities.
Finally, we would like to point out that FI reduction is the foundation of the differential equation method and AMFlow method, thus enhancing its efficiency is crucial to the whole story.

\section*{Conflict of interest}

The authors declare that they have no conflict of interest.

\section*{Acknowledgments}

The work was supported in part by the National Natural Science Foundation of China
(12325503, 11975029), the National Key Research and Development Program of China
(2020YFA0406400), the China Postdoctoral Science Foundation
(2023M733123, 2023TQ0282), and the Postdoctoral Fellowship Program of China Postdoctoral
Science Foundation.

%\bibliographystyle{elsarticle-num-names}
%\bibliographystyle{utPhysMa}
%\bibliographystyle{elsarticle-sci-bull}
%\bibliography{refs}

\end{document}